# Exploring Boolean and Non-Boolean Computing Applications of Spin Torque Devices

Kaushik Roy, Mrigank Sharad, Deliang Fan, Karthik Yogendra
Department of Electrical and Computer Engineering, Purdue University, West Lafayette, IN, USA
{kaushik, msharad, dfan, kyogendra}@.purdue.edu

**Abstract:** In this paper we discuss the potential of emerging spin-torque devices for computing applications. Recent proposals for spin-based computing schemes may be differentiated as 'all-spin' vs. hybrid, programmable vs. fixed, and, Boolean vs. non-Boolean. All-spin logic-styles may offer high area-density due to small form-factor of nano-magnetic devices. However, circuit and system-level design techniques need to be explored that leaverage the specific spin-device characterisitcs to achieve energy-efficiency, performance and reliability comparable to those of CMOS. The non-volatility of nano-magnets can be exploited in the design of energy and area-efficient programmable logic. In such logic-styles, spin-devices may play the dual-role of computing as well as memory-elements that provide field-programmability. Spin-based threshold logic design is presented as an example (dynamic resisitve threshold logic and magnetic threshold logic). Emerging spintronic phenomena may lead to ultra-low-voltage, current-mode, spin-torque switches that can offer attractive computing capabilities, beyond digital switches. Such devices may be suitable for non-Boolean data-processing applications which involve analog processing. Integration of such spin-torque devices with charge-based devices like CMOS and resistive memory can lead to highly energy-efficient information processing hardware for applicatons like pattern-matching, neuromorphic-computing, image-processing and data-conversion. Towards the end, we discuss the possibility of applying emerging spin-torque switches in the design of energy-efficient global interconnects, for future chip multiprocessors.

**1. Boolean Logic with Spin-Torque Devices:**
Recent experiments on spin-torque in device structures like lateral spin valve (LSV) [1], (fig. 1a), domain wall magnets (DWM) [2], and magnetic tunnel junctions; have opened new avenues for spin based computation. Several digital logic schemes have been proposed based on such devices. Such proposals may be classified as programmable or fixed logic styles.

**1.1 Fixed Logic Styles using Spin-Torque Switches**

All Spin Logic (ASL) proposed in [3], employs cascaded LSV's interacting through spin-torque, to realize logic gates and larger blocks like compact full adders [4], based on spin majority evaluation (fig. 1). The key feature of ASL is its compactness; however, energy inefficiency resulting from relatively larger magnet-switching delay can be identified as the down-side of it. In a standard ASL design, achieving 500MHz operation for an 8x8 multiplier would require ~60ps switching-speed for individual magnets (of size 30x15x1.5nm$^3$), leading to untenably high current-levels. This results in large static power in the ASL device (fig. 2). Non-volatility of nano-magnets can be exploited by introducing 2-phase pipelining, which would result in high-throughput for a given switching current, thereby mitigating the requirement of high-switching speed for individual magnets. Overhead due to clocking transistors can be minimized by the use of 'leaky' transistors with relatively low *ON*-resistance, without incurring significant loss in robustness [5].

The non-local STT employed in ASL decouples the biasing charge-current path from the spin-current path along which computation is performed (fig. 1). Common charge current channels can therefore be shared by multiple ASL layers in a vertical stack, leading to a high-density 3-D spin logic design. In such a design, a single CMOS substrate can be used to supply 2-phase clock to a large number of low-resistance, metallic ASL stacks, leading to high energy efficiency along with very high integration-density [5].

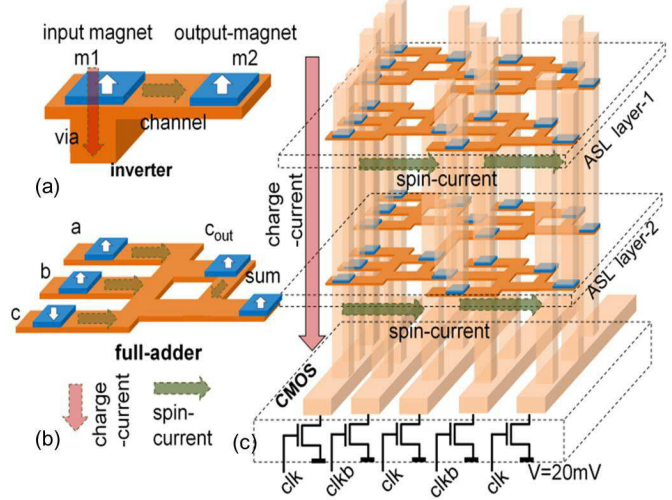

Fig. 1 (a) ASL inverter, (b) ASL full adder, (c) 3-D integration scheme for ASL with shared transistors for pipelining

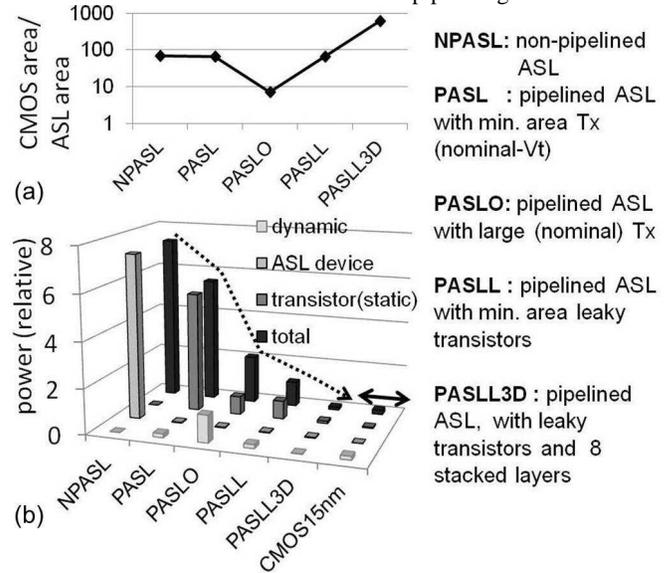

Fig. 2 (a) Area comparison of proposed ASL design with CMOS, showing the possiblity of 3 order of magnitude higher logic density, (b) different components of power dissipation in pipelined, 3D-ASL [10, 11] and their comparison with standard ASL and 15nm CMOS.

Optimized 2-phase pipelining and 3-D stacking can lead to two orders of magnitude lower power consumption as compared to a non-pipelined ASL, for a given set of device parameters. Notably, the proposed design-scheme can achieve power consumption and performance comparable to CMOS, provided desirable device parameters (like efficient magnet-channel interface, low contact resistances and high-spin diffusion length for channel) are achieved in the future [6]. The most attractive feature of 3D ASL is evidently the ultra-high area density. The prospects of achieving ~1000x higher logic density as compared to CMOS may be a motivating factor for the on-going research and experiments in this field.

Recent experiments have achieved domain wall (DW) motion with relatively small current density ($10^7$ A/cm$^2$) in magnetic nano-strips with perpendicular magnetic anisotropy (PMA) [7, 8]. Device scaling and the use of emerging spin-orbital coupling phenomena [8], can further reduce the required current density for DW-motion. Such techniques can be exploited to design ultra-low voltage logic schemes based on nano-scale domain-wall switches [9, 10]. Owing to non-volatility of DWM based logic cells, fine-grained pipelining can be achieved, leading to high throughput along with low computation energy for low and medium frequency data processing.

### 1.2. Programmable Logic Design using Spin Devices

#### A. Dynamic Resistive Threshold Logic

Non-volatile spin-torque switches can be used in designing configurable logic blocks [11-13]. Such circuits can possibly provide enhanced scalability and energy-efficiency resulting from reduced-leakage of the spin-based memory elements. Exploiting these benefits of spin-torque devices, hybrid logic circuits like magnetic full adders [12], non-volatile flip-flops [13], and memory-cells for hybrid FPGAs have been proposed [14].

Spin devices can be attractive for logic schemes that involve direct use of memory elements for computing. One such scheme is threshold logic [15]. The operation of a threshold logic gate (TLG) involves summation of weighted inputs, followed by a threshold operation as given in eq.1:

$$Y = \text{sign}(\sum In_i W_i + b_i) \quad (1)$$

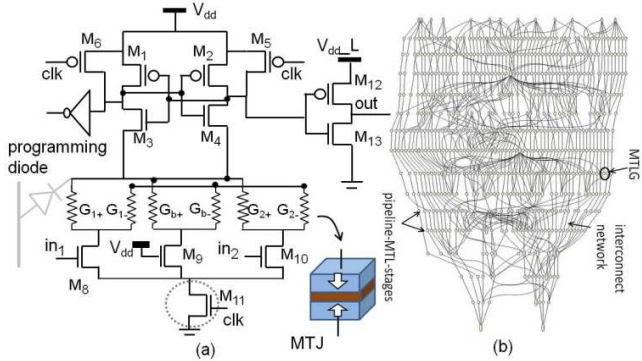

Fig. 3 (a) Magneto-resistive threshold logic gate (MTLG) with two fan-ins, (b) Pipelined threshold logic network using MTLG

Table-1 Performance Comparison: MTL vs. CMOS-LUT

| ISCAS-85 benchmark | # input | # output | delay /throughput (ns) | | Energy (fJ) | | % reduction | |
|---|---|---|---|---|---|---|---|---|
| | | | LUT | RTL | LUT | RTL | energy | energy-delay |
| c432 | 36 | 7 | 10.1 | 0.5 | 17362.56 | 480 | 97.2 | 99.86 |
| c499 | 41 | 32 | 8.18 | 0.5 | 33795.57 | 940 | 94.5 | 99.83 |
| c880 | 60 | 26 | 8.4 | 0.5 | 26394.41 | 970 | 96.3 | 99.78 |
| c1355 | 41 | 32 | 9.95 | 0.5 | 56284.24 | 1480 | 97.4 | 99.87 |
| c1908 | 33 | 25 | 11.55 | 0.5 | 56930.13 | 1200 | 97.89 | 99.91 |

Here, $In_i$, $W_i$ and $b_i$ are the inputs, weights and the thresholds respectively. A network of TLGs with configurable $W_i$ and $b_i$ can be realized with the help of spin-torque memory elements [16]. A spin-CMOS hybrid Magneto-resistive TLG (MTLG) is shown in fig. 3a. It uses magnetic tunnel junctions (MTJ) to implement the programmable input weights and the bias, connected to the input transistors ($M_8$, $M_{10}$) and a dynamic CMOS latch for the thresholding operation. Such a hybrid MTLG can be used to design pipelined, high-performance and energy-efficient TLG-array as shown in fig. 3b [16]. Table-1 compares the performance of MTLG with 4-input LUTs showing the possibility of ~100x improvement in energy-delay product as compared to CMOS-based programmable logic.

#### B. Magnetic Threshold Logic

An all-spin MTLG can also be designed that uses spin-torque switches for thresholding as well as programmable resistive weights (fig. 4). Such a spin-torque thresholding switch is shown in fig. 4a. It constitutes of a thin and short (20x60x2 nm$^3$) nano-magnet domain $d_2$ connecting two anti-parallel magnets of fixed polarity, $d_1$ and $d_3$. The domain $d_1$ forms the input port, whereas, $d_3$ is grounded. Spin-polarity of the domain $d_2$ can be written parallel to $d_1$ or $d_3$ by injecting a small current (~1µA) along it, depending upon the direction current flow [17-19]. Thus, the domain wall switch (DWS) acts as a compact, fast and low-voltage current-comparator. MTJ-based detection port is used for reading the-spin polarity of the free-domain (fig. 4). CMOS-inverter can be used to sense the state of the DWS and communicate it to the fan-out gates [20]. An all-spin MTLG can be more area efficient as compared to dynamic MTLG in fig. 3. For achieving high energy efficiency, it needs the application of small input voltages (~50mV) in order to minimize the static power consumption due to direct current-paths.

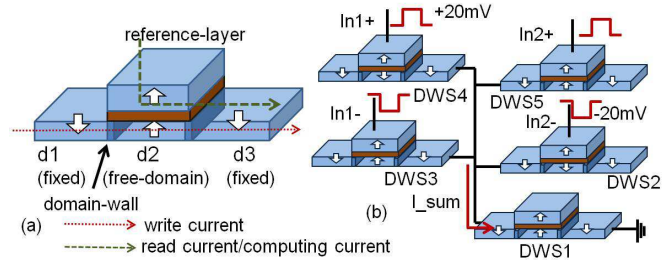

Fig. 4 (a) Domain-wall switch (DWS), (b) All-spin MTLG using spin-torque based DWS: DWS1 is the thresholding device, while DWS2-DWS4 are acting as programmable weights.

## 2. Spin-Torque Devices for Energy Efficient Non-Boolean Computing

Most of the spin based computation schemes (both hybrid as well as 'all-spin') proposed so far, have been centered on modeling digital logic gates using these devices. A wider perspective on the application of spin-torque devices, however, would involve, not only exploring possible combination of spin and charge devices but, searching for computation-models which can derive maximum benefits from such heterogeneous integration.

Ultra low voltage, current-mode operation of magneto-metallic devices like LSV's and domain-wall magnets (DWM) can be used to realize analog summation/integration and thresholding operations with the help of appropriate circuits [17-22], [29-30]. Such device-circuit co-design can lead to ultra-low power non-Boolean computation circuits. Examples of such device-circuit design are given in the following sub-sections.

### 2.1 Non-Boolean computing using spin-based circuits

Low current threshold for domain wall motion in PMA nano-magnet strips clubbed with the application of spin-orbital coupling [8], can be exploited to model a spin-torque thresholding device or 'neuron' as shown in fig 4a [17, 18]. Such a low-voltage current-mode switches can be used to for computing current-mode analog summation and thresholding operations, required in non-Boolean computing circuits like neural networks.

Fig. 5 shows neuromorphic circuit model using the DWN device described above. This example shows a 'receiving' neuron connected to two 'transmitting' neurons through synapses. A synapse is defined in terms of the magnitude and the sign of its connectivity strength or 'weight'. In the spin-based neural circuit, a dynamic CMOS latch senses the state of the DWN-MTJ while injecting only a small transient current into the detection terminal. The latch drives transistors operating in deep triode region (biased across a small drain to source voltage, ΔV~50mV), which transmit synapse-currents to all the fan-out neurons. The transistors corresponding to the positive weights, effectively source current to the receiving neurons, whereas the transistors corresponding to the negative weights act as drains. In this scheme, the spin neurons facilitate ultra-low voltage biasing of the static current-path, leading to reduced static-power consumption. Moreover, the thresholding operation of the spin-neurons can be much more energy-efficient as compared to that of conventional analog CMOS circuits [17].

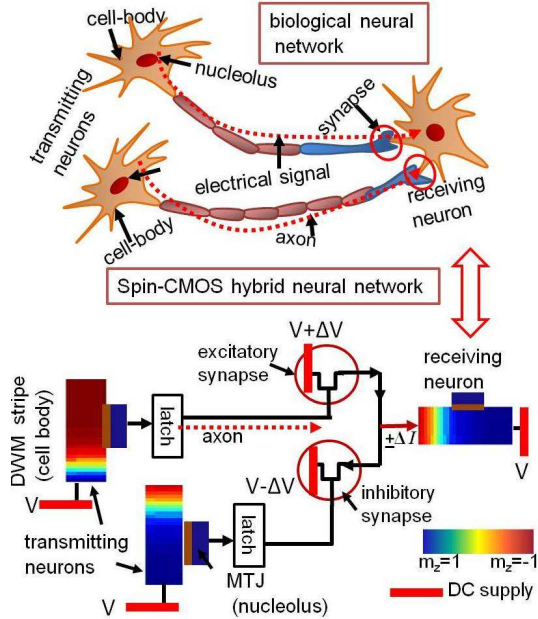

Fig. 5 Emulation of neural network using spin-CMOS hybrid circuit: In each neuron, the MTJ acts as the firing site, i.e., the nucleolus; DWM strip can be compared to cell body and its spin polarization state is analogous to electrochemical potential in the neuron cell body which affects 'firing', the CMOS detection unit can be compared to axon that transmits electrical signal to the receiving neuron, and finally a weighted transistor acts as synapse, as it determines the amount of current injected into a receiving neuron.

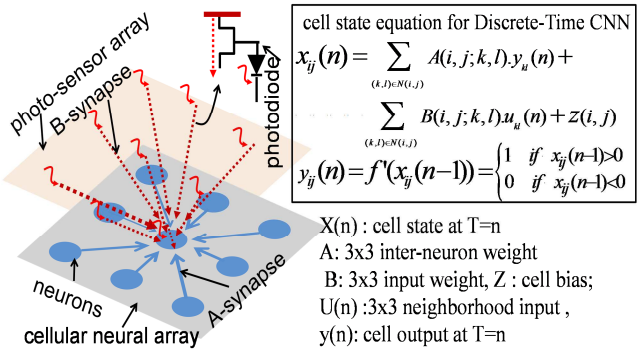

Fig. 6. 3x3 neighborhood architecture of CNN and equation for neuron's state: Current from each photosensor $u_{ij}$ is transmitted to 3x3 neighbors through type-*B* synapses implemented using weighted transistors, whereas, inter-neruron connection is determined by type-*A* synapses.

**Design Example:** The circuit concept described above can be employed in the design of different classes of neuromorphic architectures. Using this technique, we presented the design of an image processing architecture based on cellular neural network (CNN) in [30] (fig. 6). Each neuron in a CNN has two kinds of synaptic connections, type-*A* and type-*B* (fig. 6). Through the type-*A* synapses, a neuron receives the outputs $y_{ij}$, of its eight nearest neighbors and its own state as a feedback. Through type-*B* synapses, it receives the external signals, $u_{ij}$, (in this work, photo-sensor current from neighboring pixels) from 3x3 surrounding input points. The choice of the two sets of weights determines the input-output relation for the whole array and hence the image processing application. The recursive evaluation of neurons in CNN essentially involves weighted sum of these two sets of synaptic inputs, followed by a sign operation (fig. 6). In the on-sensor image processing architecture presented in [30], *A* and *B*-type synapse weights were realized using weighted triode source transistors, as described above. Note that, in this scheme, the *B*-synapse transistors receive analog-mode photo-sensor voltage at their gate, and, in turn, provide proportional currents to the neurons. On the other hand the *A*-synapse transistors receive binary voltage levels at their gates, corresponding to the source neurons' output-state.

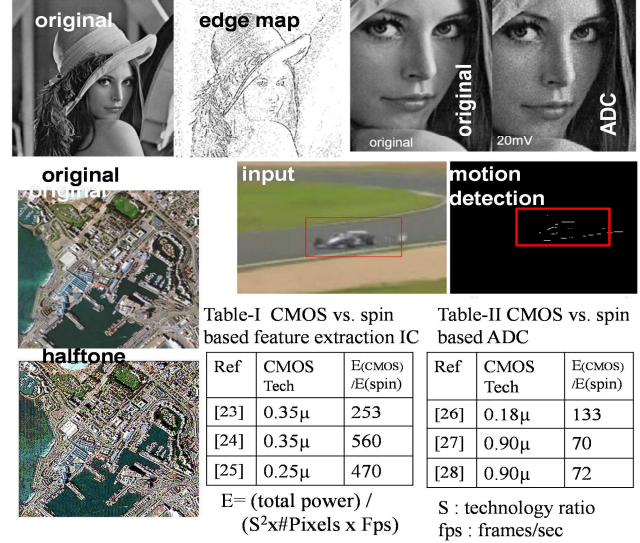

Fig. 7 Simulation results for different image processing applications: edge-extraction, motion-detection, halftoning and digitization; Table 1, 2 compares the energy per computation frame, per-pixel, of the propsoed CNN design with some recent CMOS designs for edge extraction and ADC.

Simulation results for some common image processing applications like edge-extraction, motion-detection, half-toning and digitization (fig. 7), using the spin based CNN, showed ~100x lower computation energy, as compared to state of art mixed-signal CMOS designs. As mentioned earlier, the main advantage comes from ultra low voltage, pulsed operation of spin neurons that are applied to analog computation. While subthreshold analog circuits operating at few 100mV can achieve low power [17, 30], but the corresponding switching speed is generally less than 1MHz. The spin-neurons on the other hand, while operating at ~50mV can still provide ~1GHz processing speed.

**2.2 Ultra low energy non-Boolean computing using spin neurons and resistive memory**

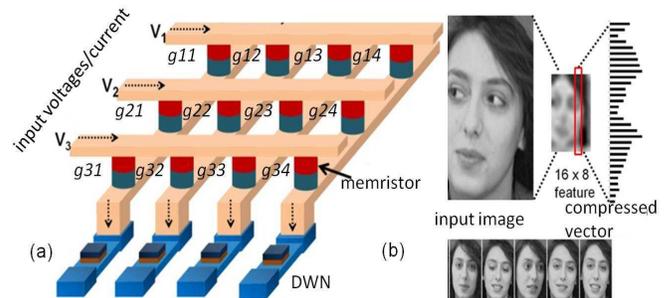

Fig. 8 (a)RCN interfaced with DWN, (b) generating of compressed vector representation of stored and input images [31]

In recent years several device-technologies have been proposed for fabricating nano-scale programmable resistive elements, generally categorized under the term 'memristor' [31, 32]. Of special interest are those which are amenable to integration with state of art CMOS technology, like memristors based on Ag-Si filaments [32]. Multi-level write techniques for memristors in crossbar arrays have been proposed and demonstrated in literature that can achieve precision up to 0.3% (equivalent to 8-bits) [32]. Such devices can be integrated into metallic crossbars to obtain high density resistive crossbar networks (RCN). The Resistive-Crossbar Network (RCN) technology has led to interesting possibilities of combining memory with computation. RCN can be highly suitable for non-Boolean computing applications that involve pattern-matching. For instance, memristors can be exploited as compact programmable weights needed in neural circuits, discussed earlier.

The RCN shown in fig. 8 constitutes of memristors (Ag-Si) with conductivity $g_{ij}$, interconnecting two sets of metal-bars ($i_{th}$ horizontal bar and $j_{th}$ in-plane bar). The horizontal bars shown in the figure receive input currents/voltages. Assuming the outward ends of the in-plane bars grounded, the current coming out of the $j_{th}$ in-plane bar can be visualized as the dot product of the inputs $V_i$ and the cross-bar conductance values $g_{ij}$ (Fig. 8). An RCN can therefore, directly evaluate the weighted summation of analog inputs and hence provides an efficient model for synapse or weighted input connections for a neural network. Each of the in-plane bars in Fig. 8 therefore can be input to an analog unit that can provide the essential neural functionality of thresholding. Several design schemes for neuromorphic hardware based on RCN have been proposed in literature that employ analog CMOS circuits to perform the thresholding task [18, 19]. As mentioned earlier, such circuits, employing current-mirrors and analog operational amplifiers (OPAMPs) lead to large static power consumption.

The critical neural functionality needed in RCN based designs can be provided by the magneto-metallic spin neurons at ultra-low energy cost. Fig. 8 depicts DWN neurons interfaced with an RCN. Since the DWN neurons are fully metallic and do not need any biasing circuitry, they facilitate the application of ultra-low input voltages. Moreover, they can provide the high-speed current-mode thresholding operation without dissipating any additional static power. Preliminary simulation results show that owing to these two factors, the spin neurons can potentially achieve more than two order of magnitude lower computation energy as compared to analog CMOS neurons. Hence, the proposed technique can facilitate the design of ultra-low energy and high performance non-Boolean computing blocks with RCN.

A design example for analog associative computing was presented in [18]. In this work, compressed representation of different face-images were stored along the in-plane columns of the RCN (fig. 8), while the dot-product between an analog input (applied to the horizontal bars) vector and the conductance values were used for detecting the best match for an input face-image. The winner detection was done with the help of spin neurons. The spin-neurons acting as compact current-mode comparators can facilitate energy-efficient successive approximation based analog to digital conversion, following which a compact and low power digital WTA can detect the highest value. Moreover, application of spin-neurons facilitates the use of ultra low input voltage levels (~50mV) in order to reduce the static power consumption in the RCN itself. Thus, ultra-low voltage, high-speed analog computing with RCN using spin neurons can achieve much higher energy-efficiency as compared to analog-CMOS circuits applied for the same task (~100X improvement). Results showed more than three order of magnitude lower computation-energy as compared to a 45nm digital CMOS design.

The basic associative unit discussed above, can be extended to a more generic architecture. The proposed design scheme can be applicable to a wide class of non-Boolean computing architectures that also include different categories of neural networks. Different spin-devices can also be used to obtain different network functionality. For, instance, long-channel DWN devices can be applicable to spiking neural networks [33].

### 3. Spin-Torque Switches for Energy-Efficient Global Interconnects

Energy-efficiency and performance of on-chip global interconnects can be a major bottleneck for emerging chip-multi-processors (CMP) that employ extensive inter-processor and memory to processor communication [36]. In [20] we proposed to application of low-voltage, magneto-metallic spin-torque (ST) switches for ultra-low energy and high-performance current-mode interconnect design. The use of CMOS-based current-mode signaling for long distance links has been shown to offer reduced power consumption and enhanced bandwidth [35]. But, analog circuits for current-mode transceivers are more complex than simple inverters, used for voltage-mode links, and add significantly to static-power consumption as well as area complexity at the I/O interfaces.

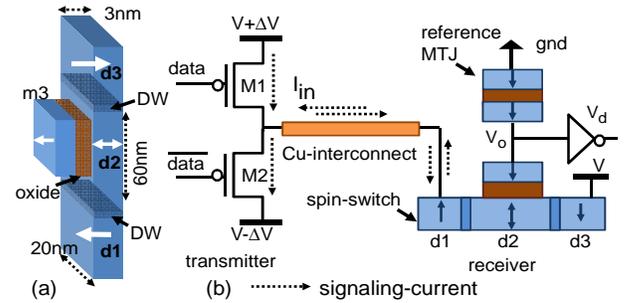

Fig. 9 (a) device structure for Domain-Wall-Switch (DWS), (b) a possible circuit for on-chip and inter-chip interconnect using DWS (variants of this device structure as well as simpler circuit configurations may be employed for interconnect design).

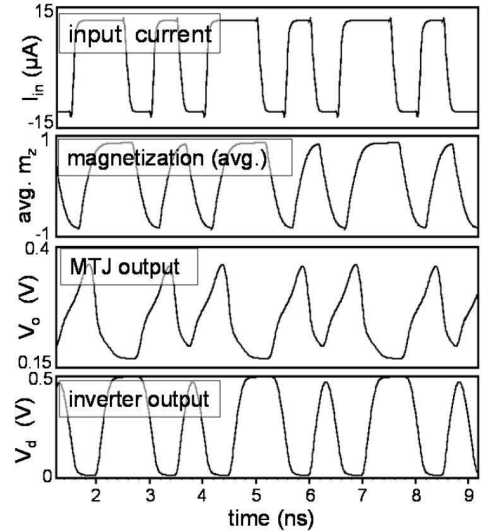

Fig. 10 (a) Simulation plots for 2Gbps signaling over 2mm long on-chip interconnect ($E_b$ =40$K_B$T, DW velocity ~200m/s for Neel-type domain-wall). (b) Switching time vs. switching current for two different anisotropy barriers, (c) power consumption in current-mode signaling and in driver and receiver circuits (including MTJs) increases linearly with signaling frequency, the oxide thickness of MTJ has been reduced for increasing frequency, in order to allow faster sensing.

Magneto-metallic spin-torque devices, like domain-wall magnet [36], and spin-valves [29, 30], can act as ideal receivers for current-mode signals, owing to their small resistance and the possibility of low-current and high-speed switching [8]. Fig. 9 depicts a possible circuit for a high-speed data interconnect employing a DWS-based receiver. At the transmitter end, the PMOS transistors M1 and M2 are driven by a voltage-mode data-signal. Their source terminals are connected to two different DC-voltages $V+\Delta V$ and $V-\Delta V$, where $V$ is 0.5V and $\Delta V$ can be ~30mV. On the receiver side, the DWS is biased at a voltage $V$, as shown in the figure. Depending upon the voltage-mode data signal at the transmitter, either $M_1$ or $M_2$ are turned *ON*, resulting in either positive or negative current flow across the DWS at the receiver. This results in data-dependent flipping of the DWS free-layer, which can be detected using a high-resistance voltage divider formed with a reference MTJ as show in fig. 9. A high TMR for the MTJ can provide a voltage swing large enough to be sensed by a simple CMOS inverter.

In the proposed interconnect-design, the energy consumption per-bit transmitted can be evaluated as the sum of the components related to static power dissipation across the transmission line $E_{int}$, and the components resulting from the power consumption in the conversion circuit $E_{conv}$, at the receiver. The dynamic switching power for the small-size digital driver at the transmitter can be negligibly small as compared to the aforementioned components.

The DWS facilitates ultra-low voltage biasing of the entire transmission link, such that the static current flows across a small terminal voltage of $2\Delta V$. A 10 mm long on-chip interconnect (parameters given in [35]) would offer a resistance of ~500Ω and an effective capacitance of ~2.5pF. Transmission of data at 2Gbps (data-period $T_d$ =0.5 ns) speed over such a link may require a current-amplitude ($I_d$) of ~15μA in order to be able to switch the DWS. This current magnitude can be supplied by minimum size transistors M1 and M2 (with effective resistance of ~1kΩ in 45nm CMOS technology) with a $\Delta V$ of ~30mV. The component $E_{int}$ can be therefore calculated as $E_{int} = 2\Delta V \times I_d \times T_d$, which evaluates to ~0.45fJ. The power consumption in the detection unit can be minimized with the optimal choice of $t_{ox}$ as discussed earlier. A TMR of 200% was used for the MTJs. For 2Gbps operation, the power consumption in the optimized detection circuit was found to be ~0.8μW. This translates to a value of ~0.4fJ for $E_{conv}$. Thus, the overall energy dissipated per-bit can be ~0.9fJ which is around two order of magnitude less than that reported in a recent mixed-signal CMOS implementation [35].

DWS switching current and hence the signaling power can be reduced by the application of spin-orbit coupling [8, 36]. This phenomena can also be conducive to high domain wall velocities of the order of ~1000m/s [38]. Such device-optimizations may facilitate more than 10GHz signaling with less than 100μA current.

The proposed spin-CMOS hybrid interconnect can be compact and area-efficient as compared to conventional mixed signal CMOS current-mode I/O interfaces. Thus the spin-torque based I/O interfaces can emerge as a very attractive solution to the design challenges associated with on-chip and inter-chip interconnects. Other spintronic switches can also be employed in the proposed scheme. For instance, low-current high-speed switching of nano-magnets in lateral spin valves may be facilitated by current-mode Bennett-clocking proposed for All Spin Logic (ASL) devices [17]. Such a device with an additional MTJ-based read-port can also be employed for high-speed, low energy current-mode interconnects. Technology cost of integrating nano-scale spin-torque switches with CMOS for interconnect-design may be comparable to that of the standard MRAM fabrication process

## 4. Conclusions

Emerging spin device phenomena have lead to interesting possibilities of low energy Boolean as well as non-Boolean computing. Non-volatility of nano-magnets can be exploited in the design of energy-efficient programmable logic hardware for hybrid FPGAs. We noted that ultra low voltage, current mode operation of magneto-metallic spin torque-devices can also be suitable for non-Boolean, analog-mode computing and global interconnect-design. Such devices may be integrated with CMOS and other charge-based devices to model energy-efficient computing systems.

**Acknowledgement:** This research was funded in part by CSPIN centre, SRC, Intel Corporation, and NSF.